\documentstyle[aps,epsf]{revtex}
\newcommand{\be}{\begin{equation}}
\newcommand{\ee}{\end{equation}}
\newcommand{\beq}{\begin{eqnarray}}
\newcommand{\eeq}{\end{eqnarray}}

\newcommand{\bm}{\boldmath}

\begin{document}

\title{The static three-quark SU(3) and four-quark SU(4) potentials}

\author{C. Alexandrou$^{1}$, Ph. de Forcrand$^{2,\>3}$, A. Tsapalis$^4$\\
$^1$ Department of Physics,
University of Cyprus, CY-1678 Nicosia, Cyprus\\
$^2$Inst. f\"ur Theoretische Physik, ETH H\"onggerberg, CH-8093 Z\"urich,
Switzerland\\
$^3$ CERN, Theory Division, CH-1211 Geneva 23, Switzerland\\
$4$ Department of Physics, University of Athens, Athens, Greece}

\maketitle

\begin{abstract}
We present results on the static three- and four-quark potentials in
SU(3) and SU(4) respectively within quenched lattice QCD.  
We use an analytic multi-hit procedure 
 for the time links  and a variational approach to  determine the ground state.
The three- and four-quark potentials extracted are consistent with a
 sum of two-body potentials, possibly with a weak many-body
component. The results  give support to the $\Delta$ ansatz for the
baryonic area law.
\end{abstract}

\section{Introduction}
The nature of the three quark potential is of prime importance in the
understanding of baryon structure. 
However up to now it
has received little attention in lattice QCD studies. This is to be
contrasted with the quark - antiquark potential relevant for
meson structure for which many lattice results exist~\cite{Bali}. 

The aim of the present work is to study the nature of the three
quark potential within lattice QCD. The fundamental question which
has been raised  more than twenty years ago, is whether the static  three
quark potential can be approximated
by a sum of three two body potentials,
known in the literature as the $\Delta$-ansatz, 
or whether it is  a genuine three body  potential. The latter is
 obtained in the strong coupling approximation by minimization
of the energy of the three quark state. The resulting 
minimal length flux tube is a configuration where the flux tubes
from each quark merge at a point. Due to its shape it is known as the
$Y$-ansatz.
 
Recently two lattice studies of the three quark potential have reached
different conclusions: Preliminary results by G. Bali~\cite{Bali} 
at $\beta=6.0$ favour the $\Delta$ ansatz whereas the
 analysis of lattice results 
at $\beta=5.7$ by Takahashi {\it et al.}~\cite{Osaka} gives more support 
to the $Y-$ ansatz. The difficulty to resolve the dominant area law
for the baryonic potential is due to the fact that the maximal
difference between the two ans\"atze
is a mere  15\%.

In our study we make a number of technical improvements in order to  
try and distinguish the $Y-$ and $\Delta-$ ans\"atze. Besides using
the standard techniques of smearing and the  multi-hit procedure 
for noise reduction,
we employ a variational approach~\cite{variational} to extract 
the ground and first excited state
of the three quarks. Both the multi-hit procedure, which is done
 analytically, and the
variational approach were not used in ref.~\cite{Osaka}.
These are  especially important for the
larger Wilson loops where the confining part of the potential is
the most dominant. Instead of the multi-hit procedure  
for the time links we have also tried the recently proposed hypercubic blocking 
\cite{hyperblocking}. We did not, however, 
find any improvement as compared to the multi-hit procedure.

In addition to the SU(3) gauge group we also present results for SU(4).
Since the same issue of which ansatz is favoured arises in any gauge group
a calculation in SU(4) can help decide the preferred
area law. The difference between the two-body approximation
and the many-body force is bigger for SU(4), reaching for the
lattice geometries that we looked at a maximum value of 20\%.

\begin{figure}
\begin{center}
\epsfxsize=10.0truecm
\epsfysize=10.truecm
\mbox{\epsfbox{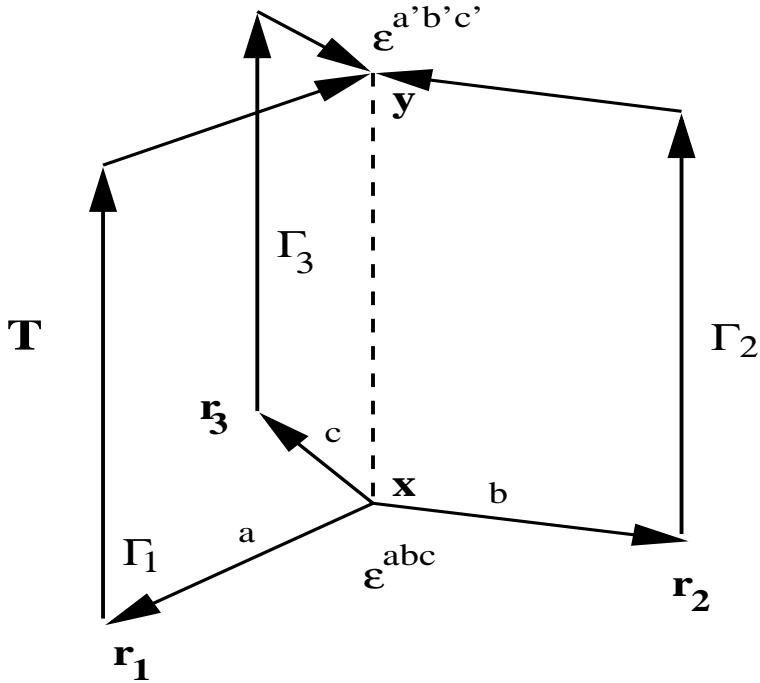}}
\end{center}
\caption{The baryonic Wilson loop. The quarks are located
at positions ${\bm r_1},{\bm r_2}$ and ${\bm r_3}$.}
\label{fig:BWL}
\end{figure}

The SU(3) baryon Wilson loop is constructed by creating a gauge invariant 
three quark state at time $t=0$ which is annihilated at a later time $T$.
\be
W_{3q}=\frac{1}{3!}\epsilon^{abc}\epsilon^{a'b'c'} U({\bm x,y},1)^{aa'}
        U({\bm x,y},2)^{bb'}U({\bm x,y},3)^{cc'}
\label{3q Wilson}
\ee
for the three quark lines that are created at $x$ and annihilated at $y$
and 
\be
U(x,y,j)=P\exp\left[ig\int_{\Gamma(j)} dx^{\mu}A_\mu(x)\right]
\ee
where $P$ is the path ordering and $\Gamma(j)$ denotes the path from $x$ 
to $y$ for quark line $j$ as shown in Fig.~\ref{fig:BWL}.

The three quark potential is then extracted
in the standard way from the long time behaviour of the Wilson loop:
\be
V_{3q}=-\lim_{T \rightarrow \infty} \frac{1}{T} \ln<W_{3q}> \quad.
\label{3q pot}
\ee

In SU(4) the corresponding colour singlet gauge invariant four
quark state is constructed in an analogous manner
and the four quark potential $V_{4q}$ is
similarly extracted. We will be using the term baryonic potential
to denote the colour singlet combination of $N$ quarks despite the fact
that in SU(4) the spin of the four quark state is an integer.

\section{Wilson loop for three and four quarks}
\subsection{SU(3)}
We describe in more detail here the two possibilities 
put forward for the area-law
of the  SU(3) baryon Wilson loop. 
In the strong
coupling limit in the presence of three heavy quarks the
gauge  invariant three quark
 state with the least amount of flux will yield the lowest
energy.  
If the three quarks are at positions ${\bf r_1},{\bf r_2}$ and
${\bf r_3}$ and
provided none of the interior angles of the triangle with vertices at
the quark sites is greater than $120^{0}$
then the flux tubes from the quarks  will meet at an interior point
${\bm r_4}$~\cite{CKP}. The position ${\bm r_4}$ is determined by minimizing
the static energy with the result
\be
\sum_{k=1}^3 \frac{ ({\bm r_k}-{\bm r_4})}{|{\bm r_k}-{\bm r_4}|} = 0 \quad,
\label{Steiner point}
\ee
which is known as the Steiner point. The angles between the flux tubes 
are $120^{0}$
independently of the vectors ${\bm r}_k$. If one of the interior angles
 of the triangle of the quarks is greater than $120^{0}$ then the flux tube
at that angle collapses to a point. Time evolution of this state produces 
a three-bladed area.
This area law is the Y-ansatz mentioned in the Introduction.
We denote the minimal
length of the flux tube for this ansatz $L_Y$ and the corresponding area $A_Y$.

The second possibility~\cite{Cornwall} is that the relevant area 
dependence of the baryonic Wilson loop is given by the sum
of the minimal areas $A_{ij}$ spanning quark lines $i$ and $j$.
This is known as the $\Delta$-ansatz with the corresponding
length and area denoted by $L_\Delta$ and $A_\Delta$
respectively.

The position of the Steiner point can be obtained analytically~\cite{CKP}
in terms of the three quark positions and the difference between
the two laws as compared to the two-body ansatz, 
\be
\biggl(\sum_j r_{j4} - \frac{1}{2}\sum_{j<k} r_{jk}\biggr)/
        \frac{1}{2}\sum_{j<k}r_{jk} \quad,
\label{Y-Delta ratio}
\ee
attains~\cite{CKP} the maximum 
value of $(L_Y-L_\Delta/2)/(L_\Delta/2)=2/sqrt{3}-1=0.154..$ 
when the quarks form an equilateral triangle. 
The factor of 1/2 is due to the non-Abelian nature of the gauge
couplings giving half as much attraction for a $qq$ 
in an antisymmetric colour state
as a $q\bar{q}$  in a colour singlet. In general the attraction
for  $(N-1)$ quarks in an $N$ quark antisymmetric colour state is a factor
$1/(N-1)$ less that the attraction for a $q\bar{q}$
in a colour singlet.
Because of this factor $ L_\Delta/(N_1) \leq L_Y$.

\subsection{SU(4)}

\begin{figure}
\begin{center}
\epsfxsize=9.0truecm
\epsfysize=8.5truecm
\mbox{\epsfbox{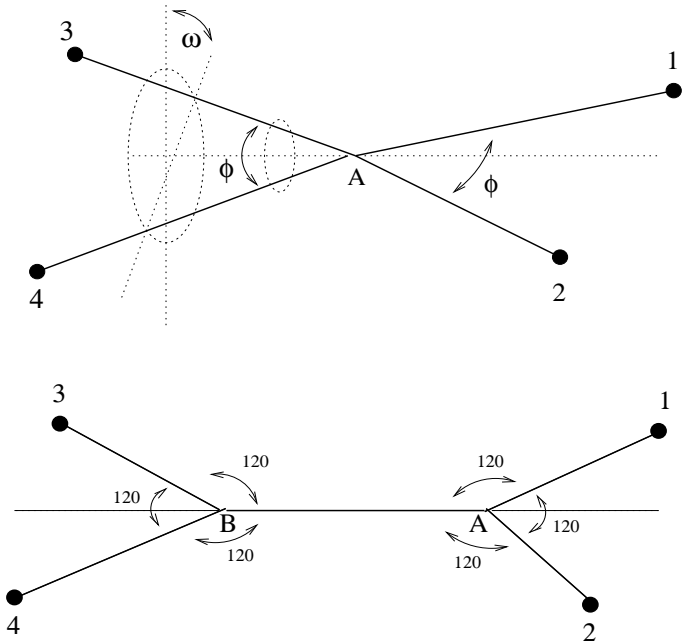}}
\end{center}
\vspace*{0.5cm}
\caption{The Wilson loop for four quarks. The quarks are located
at positions ${\bm r_1},{\bm r_2}$, ${\bm r_3}$ and ${\bm r_4}$. 
The upper graph shows the local minimum of the energy
 with one Steiner point A, and
the lower is the minimum with two Steiner points A and B.}
\label{fig:BWL su4}
\end{figure}

In SU(4)
the ground state of the system in strong coupling 
corresponds to the configuration with 
minimal length for the flux tubes which join the quarks.
Minimization of the static energy results 
in the introduction of two Steiner points, A and B
somewhere in space, with the flux tubes from two quarks joining at A,
while the flux tubes from the other two quarks meet at B. This configuration
is visualized in Fig.~\ref{fig:BWL su4}.
Since $4\times 4= 6 \bigoplus 10$  the two lines emanating from the
 two Steiner points join to form a colour singlet. 
In analogy to SU(3) we will
 call this area law  as the $Y$-ansatz, with a corresponding
flux tube length $L_{Y}$.

The two Steiner configuration is always favoured as compared to a
single Steiner point defined by equation
\be
\sum_{i=1}^4 \frac{ ({\bm r_i}-{\bm r_A})}{|{\bm r_i}-{\bm r_A}|} = 0 \quad .
\label{Steiner X}
\ee
 where all four quark lines meet 
and  which is a local minimum of the static energy.
The area law for the baryonic Wilson loop takes
now the shape of a long four-bladed surface with the blades meeting at $A$
as shown in fig.~\ref{fig:BWL su4}.
Due to this shape, we refer to this configuration as the $X$-law and 
denote the corresponding flux tube length as $L_X$.

In contrast to SU(3) where
 for any given location of the three quarks, the Steiner
point and therefore the energy  can be computed analytically,
in $SU(4)$, the two Steiner points in the $Y$-ansatz
can be obtained by a simple iterative numerical procedure. 
The two Steiner points have vectors that meet each at $120^{0}$
and one Steiner point can be obtained in terms of the other.
Starting from  an initial guess for the position of one of the Steiner
points, ${\bm r}_A$, we can compute 
$r_B$ as the Steiner point of ${\bm r}_3$, ${\bm r}_4$,
${\bm r}_A$. The ${\bm r}_1$,${\bm r}_2$, ${\bm r}_B$ vectors
lead now to a new estimate for a Steiner point ${\bm r}_A$ which in turn
is used to compute a new ${\bm r}_B$ etc. 
[The procedure converges after 30-40 iterations to the minimum.]
The location of the 
single Steiner point is easily computed by a
numerical solution to eq.~\ref{Steiner X}. 

It has been argued in ~\cite{Cornwall} that the two-body force is the 
relevant interaction for any $SU(N)$ gauge theory. It is proven in 
~\cite{Cornwall} that $L_{Y} \ge L_\Delta /(N-1) $ holds for any
location of the four quarks. From the numerical investigation , it
turns out that the relative difference between the $Y-$ energy and the two-body law
is maximal for the configuration of maximal symmetry for the four quarks.
This amounts to putting the quarks on the vertices of the regular
tetrahedron and gives a relative difference of $21.96 \,\, \% $ with respect to the 
two-body term.
This difference decreases as the configuration becomes more asymmetric in 
space and can decrease down to $5-6 \,\, \% $ for the most asymmetric 
locations of the quarks on a $16^3$ lattice.
Therefore, in order to obtain a clear signal on which law is preferred by the
$SU(4)$ quarks, we studied geometries with maximal symmetry. 

As far as the four-bladed surface area law is concerned, we observed that
$L_X$ 
always exceeds  $L_{Y}$  by at most $3.5 \,\, \%$.
In fact, the ratio, $(L_X - L_{Y}) / L_{Y} $,
becomes minimal for the most symmetric configuration of the tetrahedron, 
obtaining a value of just $0.43 \,\, \%$.
Here the ratio  in fact increases as one
increases the asymmetry of the four quark locations,
becoming maximal if all four quarks are located on
a plane. In particular, if the quarks are located at the vertices of
a square, $(L_X - L_{Y}) / L_{Y} $ takes its maximal value of $3.5 \,\, \%$. 
With the current data, discriminating an effect of ${\cal O}(3)\,\, \%$ between the
$Y$- ansatz and the $X-$ ansatz  is not possible. Therefore, we will pick
geometries that maximize the difference between the $Y$- and $\Delta$ 
ans\"atze.
Since for these geometries the difference between the $Y$- and $X$-ans\"atze 
is of the order of $2\%$ 
one has to keep in mind that when we refer to  
the $Y$-ansatz  we will in fact mean the area law with 
one or two Steiner points.

The factor of $1/(N-1)$ which relates the long range part of the 
two-body $q\bar{q}$ and $qq$ potentials
 also occurs in lowest order gluon exchange so that the two-body short range
potential is given by~\cite{Cornwall}   
\be
\frac{1}{(N-1)} \sum_{j<k} V_{jk} \quad,
\label{N coulomb}
\ee
 where
$V_{kj}$ is the $q\bar{q}$ one-gluon potential
\be
V_{jk} = -\frac{g^2 C_F}{4\pi r_{jk}}
\label{Coulomb}
\ee
with $C_F=(N^2-1)/2N$ the quark Casimir of O(N).   
 
Thus the expected forms of the ``baryonic'' potential in SU(N) that
we will be applying to SU(3) and SU(4) are
\be
V_{Nq}({\bm r_1,\cdots,r_N}) = \frac{N}{2} \!V_0 
       -\frac{1}{N-1}\sum_{j<k}\frac{g^2 C_F}{4\pi r_{jk}} 
+ \frac{1}{N-1}\sigma L_\Delta 
\label{Delta ansatz}
\ee
or
\be
V_{Nq}({\bm r_1,\cdots,r_N}) = \frac{N}{2} \!V_0 
       -\frac{1}{N-1}\sum_{j<k}\frac{g^2 C_F}{4\pi r_{jk}} 
+ \sigma L_Y 
\label{Y ansatz}
\ee
with $\sigma$ the string tension of the $q\bar{q}$ potential.
 
\section{Lattice techniques}
As we have mentioned in the Introduction, the
two recent lattice studies of the baryonic potential
~\cite{Osaka,Bali} have yielded different conclusions,
the first supporting the $Y-$ ansatz and the second the $\Delta-$ ansatz.
Since the difference between the 
two ans\"atze is $\sim 15\%$ for SU(3), obtaining conclusive results requires
making a large effort to reduce the statistical noise, especially for the large loops
where the absolute difference between the two ans\"atze becomes more visible.
In this work, we used a number of improvements as compared to previous
studies in SU(3). To our knowledge, this is the first measurement of the 4-quark potential 
in SU(4).
We describe briefly the techniques that we use in order to reduce noise
and extract more reliably the ground state.
\begin{itemize}
\item We use the multi-hit procedure~\cite{multihit} for the time links.
For SU(3) the temporal links are integrated out analytically\cite{onelink} and substituted by
their average value
\be
U_4(x) \rightarrow \bar{U}_4(x) =\frac{\int dU \> U_4(n) \>e^{\beta S_4(U)}}
{\int dU  \>e^{\beta S_4(U)}}
\label{multihit}
\ee
with $S_4(U)=\frac{1}{N} {\rm Tr}(U_4(n)F^\dagger(n))$ and $F(n)$ is the
staple attached to the time 
link that is being integrated over.
It has been 
shown in SU(2)~\cite{multihit} that replacing the time
links by their average value in this fashion reduces
the error on large Wilson loops of the order of tenfold.
The factor found in ref.~\cite{multihit} is $x^{2T}\sim 0.889^{2T}$ where
$T$ is the time extent of the Wilson loop.
For the SU(3) baryon loop the reduction factor will be
 $x^{3T}$ giving an even larger noise reduction for the large loops.
 We point out here
that the multi-hit procedure was not used in ref.~\cite{Osaka}.
In SU(4) the integration over the temporal links was done numerically. 

\item We compared the multi-hit procedure with
 the recently proposed hypercubic blocking~\cite{hyperblocking} 
on the time links. Using the optimal parameters given in ~\cite{hyperblocking}
at $\beta=6.0$,
we compare in Fig.~\ref{fig:hyperblocking} the results
on the same configurations, using the analytic multi-hit procedure 
and using hypercubic blocking.
As it can be seen, the multi-hit procedure gives smaller
errors for large loops and therefore we adopt it in this work.

\begin{figure}
\epsfxsize=9.0truecm
\epsfysize=9.truecm
\mbox{\epsfbox{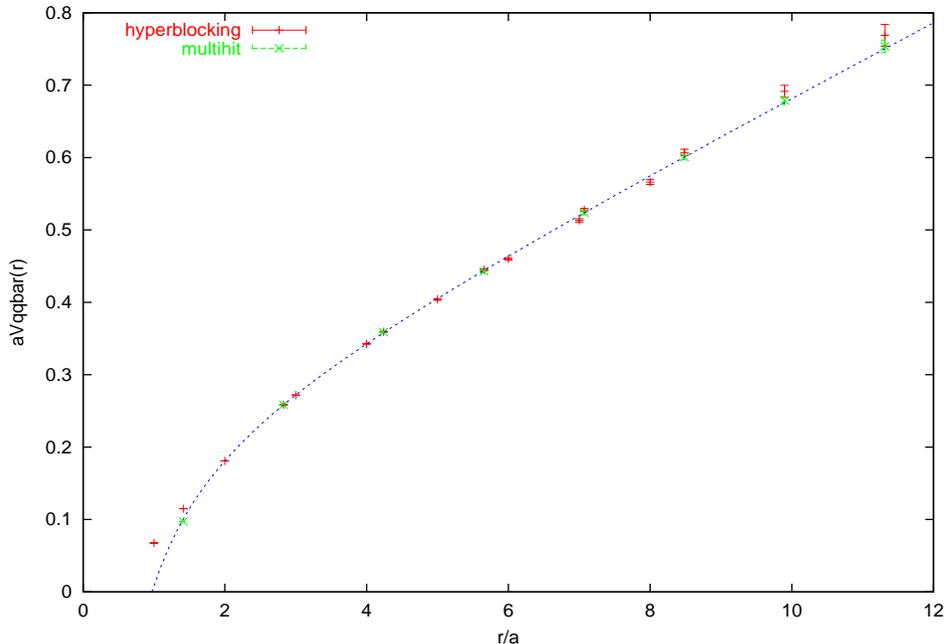}}
\caption{The $q\bar{q}$ potential for SU(3) at $\beta=6.0$ on a $16^3\times32$ 
lattice using the multi-hit procedure with x-symbol and hypercubic blocking shown
with the crosses.}
\label{fig:hyperblocking}
\end{figure}

\item To maximize the overlap of the trial state
with the three quark ground state  we use smearing of the 
spatial links~\cite{APE}.  We replace each spatial link by
a fat link by acting on it with the smearing operator ${\cal{S}}$ 
defined by
\be
{\cal{S}}{U}_j(x) = {\cal{P}}\Biggl(U_j(x) + \alpha\sum_{k\neq j}\left[
    U_k(x)U_j(x+a\hat{k})U_k^\dagger(x+a\hat{j})\right]\Biggl) \quad,  
\label{APE snearing}
\ee
where ${\cal{P}}$ denotes the projection onto SU(3). This is iterated
$n$ times.
We consider $M$ different levels of smearing and construct an $M\times M$
correlation matrix of Wilson loops~\cite{ALPHA}. For the parameter $\alpha$
and the  number of
smearings, $n_l$, for each different
smearing level $l$ we   take what is
found to be optimal in~\cite{ALPHA}; namely
\be
\alpha=\frac{1}{2} \hspace*{2cm} 
n_l\approx \frac{l}{2}\left(\frac{r_0}{a}\right)^2
\ee
for smearing levels $l=0,\cdots, M-1$ and $r_0$ Sommer's reference 
scale~\cite{Sommer}.
In all our  computations we used $M=4$. For SU(4) at $\beta=10.9$ we found
that the parameters used for $\beta=5.8$ in SU(3) produce reasonable
results.

The correlation matrices $C(t)$ for the mesonic and baryonic Wilson loops were
analyzed using a variational method~\cite{variational}. We use
two different variants both yielding consistent results.

In both variants we solve the generalized eigenvalue problem~\cite{ALPHA}
\be
C(t)v_k(t)=\lambda_k(t)C(t_0)v_k(t)
\label{eigenvalues}
\ee
taking $t_0/a=1$. 
In the first variant, the potential levels are  extracted via
\be
aV_k={\rm Lim}_{t\rightarrow \infty} -\ln
\left(\frac{\lambda_k(t+1)}{\lambda_k(t)}\right)
\label{ratio of eigen}
\ee  
by fitting to the plateau.  
In the second variant we consider the projected Wilson loops 
\be
W_P(t)=v_0^T(t_0)C(t)v_0(t_0)
\label{projection}
\ee
and fit to the plateau value of  $-{\rm ln}\biggl(W_P(t+1)/W_P(t)\biggr)$.
In Fig.~\ref{fig:Ph_Dina} we show the results of these two variants
for SU(4) for the four quark static potential.

\begin{figure}[hp]
\begin{center}
\epsfxsize=15.0truecm
\epsfysize=11.truecm
\mbox{\epsfbox{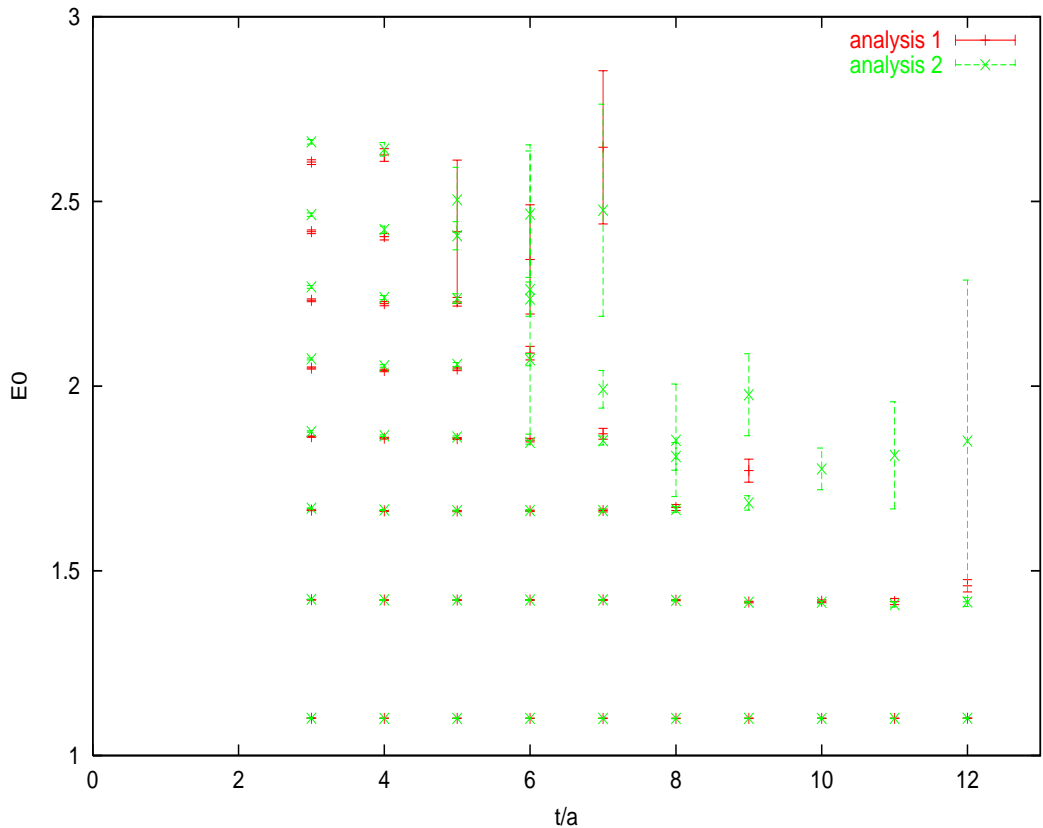}}
\end{center}
\caption{Comparison of the plateaus obtained in SU(4) for 
$-{\rm ln}(\lambda_0(t+1)/\lambda_0(t)$ solving 
the generalized eigenvalue equation
at each time $t$ (pluses) and with the projected Wilson loop
$-{\rm ln}\biggl(W_P(t+1)/W_P(t)\biggr)$ (x-symbols).}  
\label{fig:Ph_Dina}
\end{figure}

The projected correlation has a larger
contamination of excited states for time slice $t/a=3$ but 
by the next time slice the two procedures yield the same results.
We have found that  for SU(3), where the $n_l$
for the $l$-th smearing level are larger for the corresponding $\beta$ 
values than the number of smearings used in SU(4),
the projected method yields 
smaller errors. In all cases we have checked that the values
we extract for the ground state within these two procedure are consistent
with each other.
From Eq.(~\ref{ratio of eigen}) we also obtain the energy for the
first excited state. Although the data are rather noisy, we can obtain
an estimate, which we use to fix the minimum value for the time interval
 used
in the extraction of the ground state, such that the
excited state contamination is less than $e^{-2}$. 

\end{itemize}

\section{Results}

All the computations were carried out on  lattices of size $16^3\times 32$
at $\beta$ values $5.8$ and $6.0$ for SU(3) and $10.9$ for SU(4). The
$\beta$ value
 for SU(4) was chosen so that the lattice spacing is close to the value
for SU(3) at $\beta=6.0$. In the case of SU(3) we used 200 configurations
at $\beta=5.8$ and 220 at $\beta=6.0$
available at the NERSC archive~\cite{connection} and for SU(4) we generated 100
quenched configurations. 

\begin{figure}
\epsfxsize=10.0truecm
\epsfysize=10.truecm
\mbox{\epsfbox{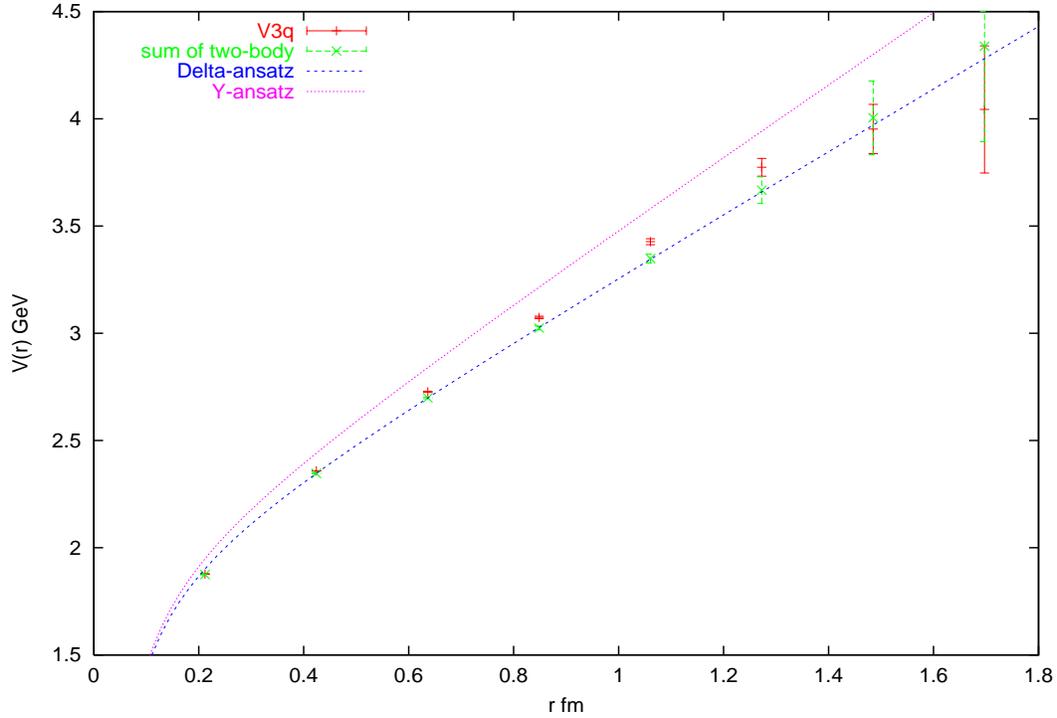}}
\caption{The static baryonic potential at $\beta=5.8$ (pluses) in physical
units. The crosses
is the sum of the static $q\bar{q}$ potential. The curves for the 
$\Delta$ and $Y$ ans\"atze are also displayed. The quarks are located
at $(l,0,0)$, $(0,l,0)$, $(0,0,l)$ and $r = r_{12} = r_{13} = r_{23} = 
\sqrt{2} l \,\, $.}
\label{fig:beta58}
\end{figure}

\begin{figure}
\epsfxsize=10.0truecm
\epsfysize=10.truecm
\mbox{\epsfbox{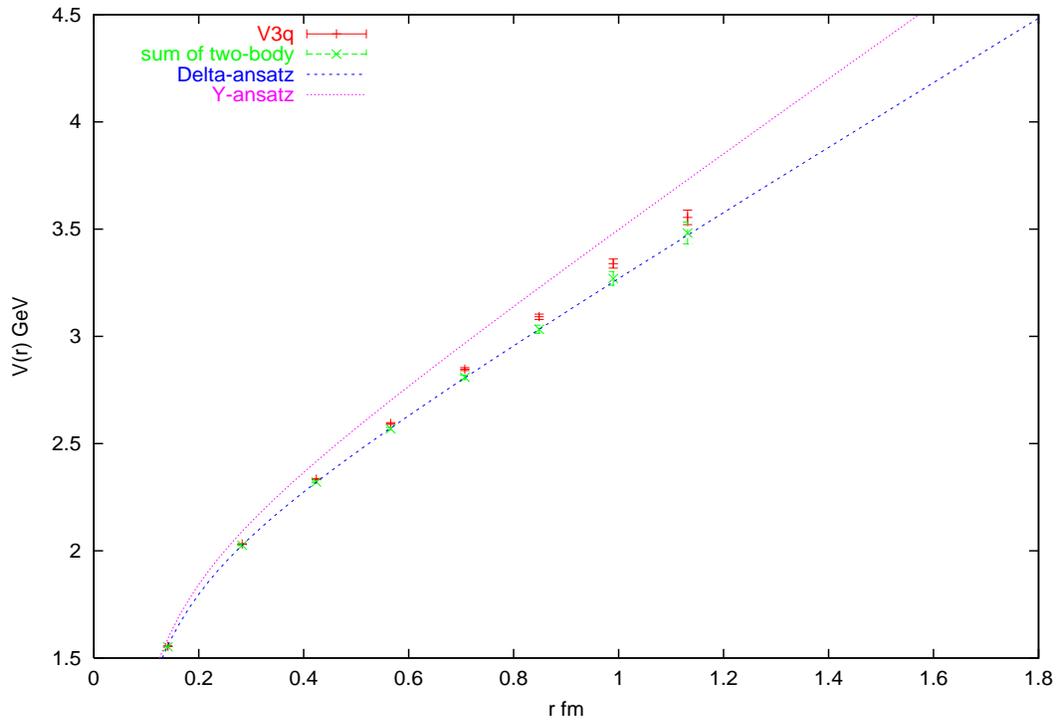}}
\caption{As figure~\ref{fig:beta58} but for $\beta=6.0$. }
\label{fig:beta60}
\end{figure}

We consider geometries on the lattice  which produce
 the biggest difference between
the $\Delta-$ and $Y-$ ans\"atze. For SU(3)
each quark is placed on a different
spatial axis equidistant from the origin.
The results are shown in Figs.~\ref{fig:beta58} and
\ref{fig:beta60} for $\beta=5.8$ and 6.0 respectively.
 To reduce systematic errors when comparing 
with the $q\bar{q}$ potential, we also compute, on the same configurations, the static $q\bar{q}$ potential with
the quark and the antiquark at the same locations as the 3 quarks of the $qqq$ potential.
The errors shown on these figures are the jackknife errors.
The string tension in lattice units
extracted from fitting the $q\bar{q}$ potentials
 is $a\sqrt{\sigma}=0.329(3)$ at $\beta=5.8$ and $0.224(3)$ at $\beta=6.0$
 consistent with the value of ref.~\cite{Bali2}. 
 At short distances
the baryonic potential, $V_{3q}$, is approximately equal to the sum of the
corresponding two - body potentials i.e. we find agreement with the tree 
level result that $V_{3q}\approx 3/2 V_{q\bar{q}}$. At larger distances,
$V_{3q}$ is enhanced compared to the tree level result. On the same figures
we also show the curves corresponding to the $\Delta-$ and $Y-$ ans\"atze. 
The 
lattice data lie closer to the curve given by the $\Delta-$
area law. However, at distances larger than about 0.7~fm,
the three-quark potential appears enhanced as compared to
 the sum of the two-body potentials. This enhancement can be explained by
a small admixture of a three body-force, although it is is so small that it
might also reflect imperfections in our variational search for the groundstate.

\begin{figure}
\epsfxsize=10.0truecm
\epsfysize=10.truecm
\mbox{\epsfbox{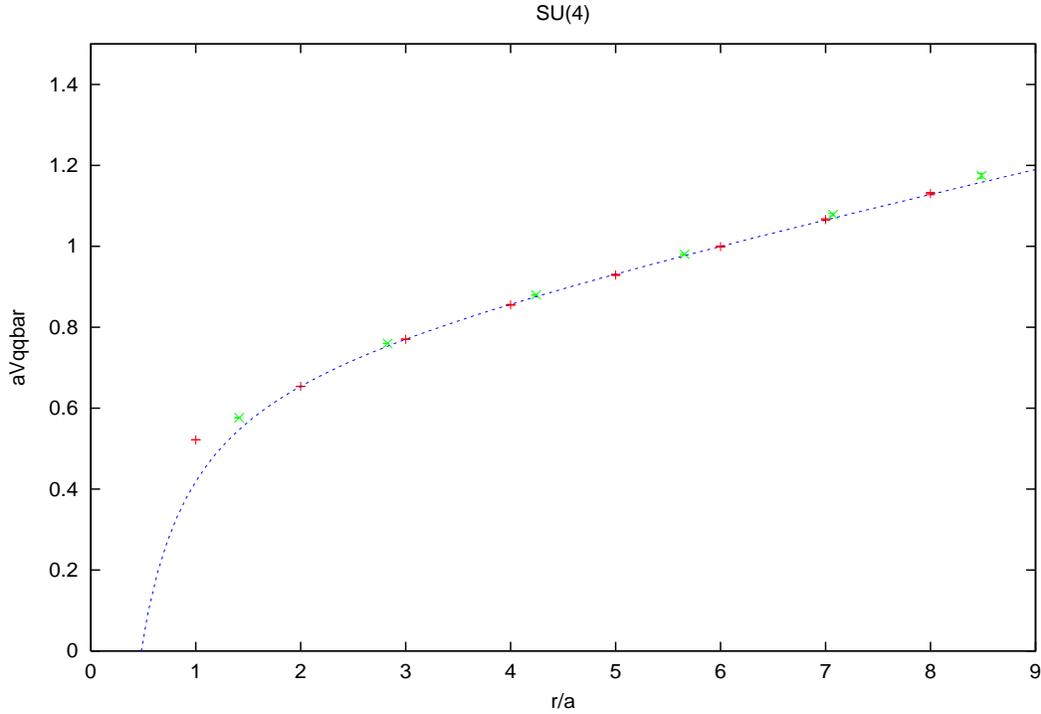}}
\caption{The $q\bar{q}$ potential for SU(4) at $\beta=10.9$ fitted to
the form $V_0-b/r+\sigma r$. The errors shown are the jackknife errors.}
\label{fig:Vqqbar su4}
\end{figure}

In SU(4) we studied three different geometries chosen so that the
difference between the $\Delta$ and analog of the $Y$ law is maximal.
In what we call geometry 1  the  quarks are placed symmetrically on  a plane
distance $l$ from the origin. 
The energy difference between the two ans\"atze is $20.0 \%$. In
geometries 2 and 3,
 three quarks have coordinates (l,0,0),(0,l,0), (0,0,l) whereas the fourth 
is at (0,0,-l) for geometry 2 and at the origin for   geometry 3.
The energy differences between the $\Delta$ and
$Y$ laws are $20.1 \%$ for geometry
2 and $19,1 \%$ for geometry 3.
The string tension  is obtained
 by fitting the on axis $q\bar{q}$ potential excluding the first point. 
We find $a\sigma=0.238(4)$ in agreement with  the  value of 
$0.2429(14)$ of ref.~\cite{Teper}.
The quality of the fit is shown in
Fig.~\ref{fig:Vqqbar su4} with $\chi^2/{\rm d.o.f}=1.0$ where we included
the results when the quark and the antiquark are on different axes.

\newpage

\begin{figure}[p]
\epsfxsize=10.0truecm
\epsfysize=7.0truecm
\mbox{\epsfbox{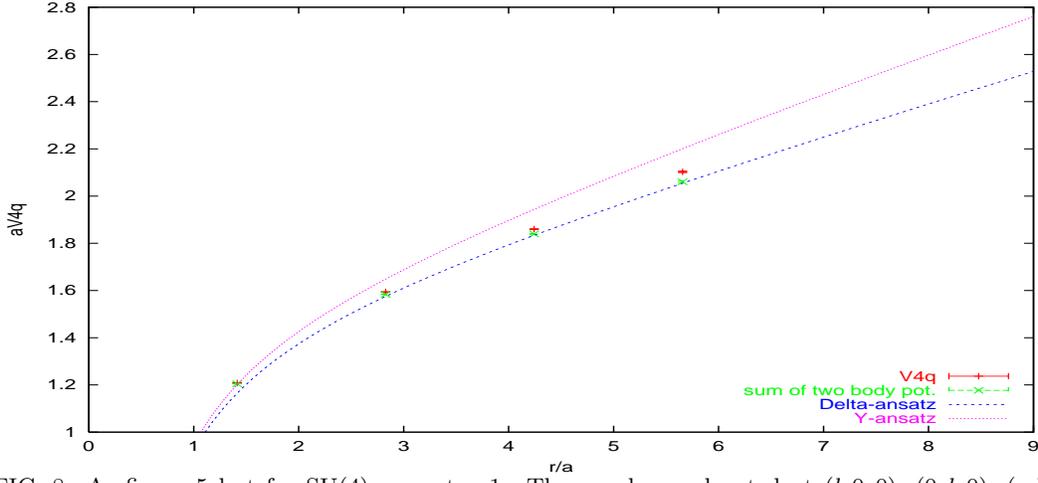}}
\caption{As figure~\ref{fig:beta58} but for SU(4) geometry 1. The quarks are
located at $(l,0,0)$, $(0,l,0)$, $(-l,0,0)$, $(0,-l,0)$ and  $r = r_{12} =
r_{23} = r_{34} = r_{14} = \sqrt{2} \,\, l $. }
\label{fig:su4 geometry 1}
\end{figure}

\begin{figure}[p]
\epsfxsize=10.0truecm
\epsfysize=7.0truecm
\mbox{\epsfbox{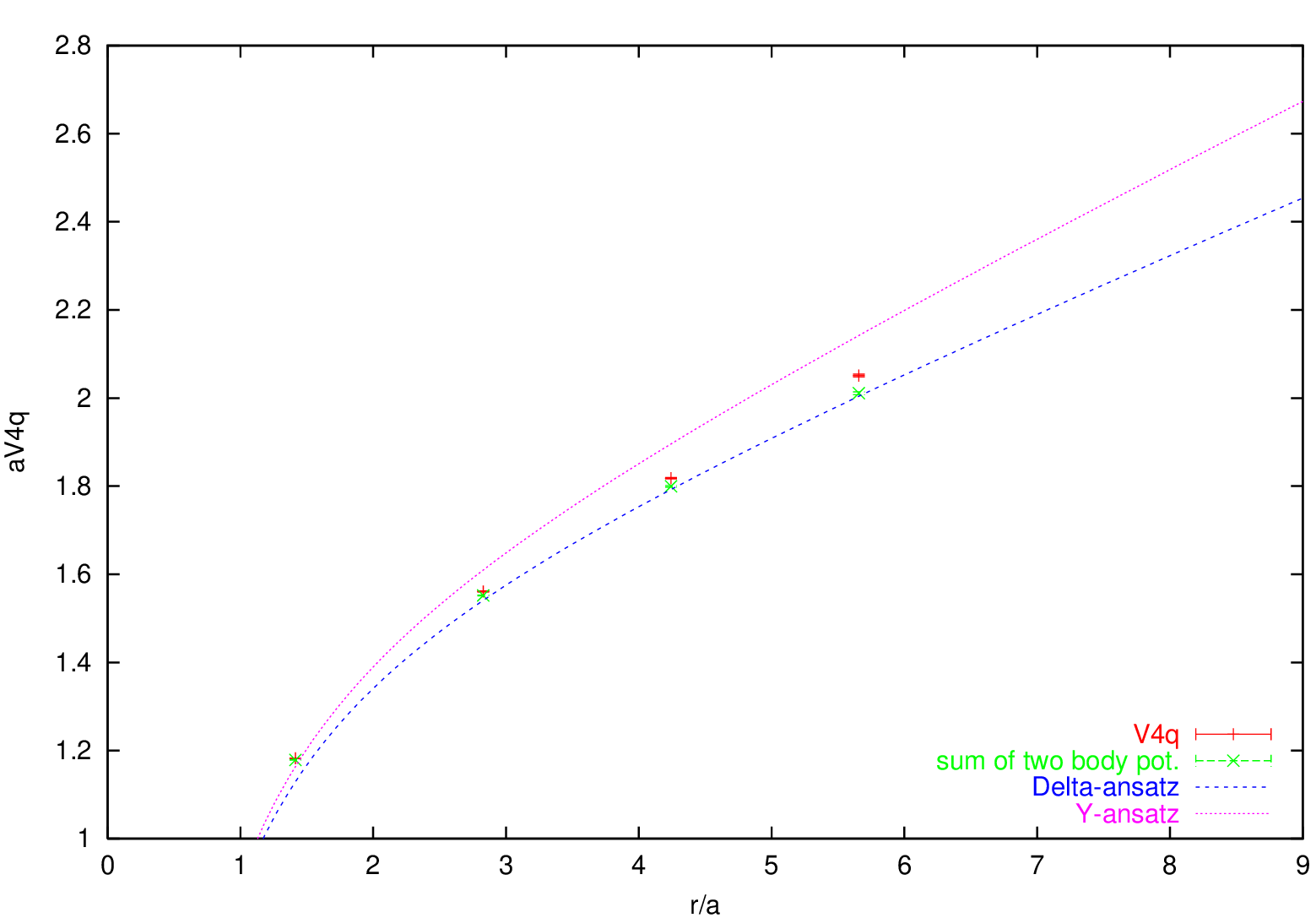}}
\caption{As figure~\ref{fig:beta58} but for SU(4) geometry 2. Here the quarks
are located at $(l,0,0)$, $(0,l,0)$, $(-l,0,0)$, $(0,0,l)$ and $r = r_{12} = 
r_{23} = r_{34} = r_{14} = \sqrt{2} \,\,l $. }
\label{fig:su4 geometry 2}
\end{figure}

\begin{figure}[p]
\epsfxsize=10.0truecm
\epsfysize=7.0truecm
\mbox{\epsfbox{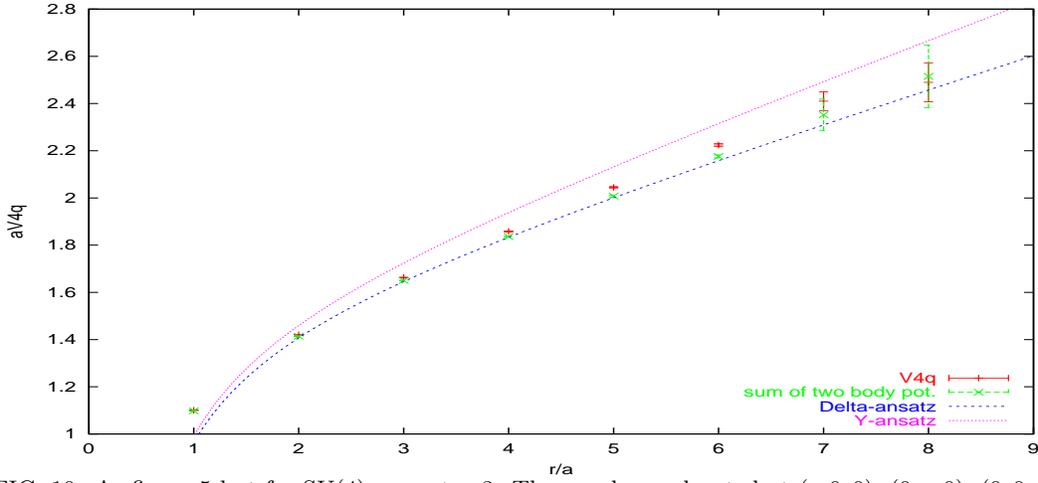}}
\caption{As  figure~\ref{fig:beta58} but for SU(4) geometry 3. The quarks
are located at $(r,0,0)$, $(0,r,0)$, $(0,0,r)$ and $(0,0,0)$.}
\label{fig:su4 geometry 3}
\end{figure}

The corresponding results for the four-quark static potential
 are shown in Figs.~\ref{fig:su4 geometry 1}, ~\ref{fig:su4 geometry 2} 
and \ref{fig:su4 geometry 3} for geometries 1, 2 and 3 respectively. 
Again we find that the four quark potential is approximated by the sum of
$q\bar{q}$ potentials with a small enhancement at larger distances. 
The results in all cases lie closer to the $\Delta$ ansatz.

\newpage
\vspace*{-2cm}
\vspace*{-1.0cm}
\vspace*{-1.0cm}

\section{Conclusions} 
Our results for the static three- and four-quark potential
in SU(3) and SU(4) are consistent with the sum of two-body potentials
below a distance of about $0.8$~fm, and clearly inconsistent with the $Y-$ ansatz. 

For larger distances, where our statistical and systematic errors both
become appreciable, there appears to be a small enhancement due to
an admixture of a many-body component. 
Nevertheless, for the distances
up to 1.2~fm that we were able to  probe in this work, 
the $\Delta$ area law gives the closest description of our data. 

We have made use of all the known techniques in order to reliably identify
the plateaus in the Wilson loops and extract the ground state potential.
Nevertheless, for the larger loops the plateaus were hard to identify,
resulting in large errors. This is a challenging numerical problem, and we cannot 
exclude the possibility that the small enhancement of the potential above the $\Delta$ area law 
which we observe is simply caused by a failure to filter out all 
excited states in our variational
search for the groundstate.
Taking the results in  both SU(3) and SU(4) at face value
the conclusion that can be drawn is that the $\Delta$ area law 
provides the closest
description to the baryonic potential up to distances of $1.2$~fm.
More refined techniques for noise reduction 
for the large loops will be needed
in order to clarify whether a genuine many-body component is present
at larger distances.

\vspace*{1cm}

\noindent
{\bf Acknowledgments:} We thank E. Follana and H. Panagopoulos
 for discussions. A. T. wishes
to thank the University of Cyprus for extended hospitality and 
financial support during stages of this research.

\end{document}